# Life-cycle structural reliability of RC bridge piers under corrosion in a changing climate

G.V. Nava, L. Capacci & F. Biondini
*Politecnico di Milano, Milan, Italy*

L. Casti
*Université Gustave Eiffel, Champs-sur-Marne, France*

ABSTRACT: Aging and deterioration processes of structures and infrastructure systems can be severely affected by the long-term evolution of environmental conditions expected in a changing climate. This paper aims at investigating the interaction between climate change and environmental aggressiveness on the probabilistic life-cycle performance of concrete structures under corrosion. A simulation-based framework is proposed for life-cycle structural reliability analysis under climate change scenarios. The framework is applied to a bridge pier with hollow core cross-section exposed to chloride-induced corrosion. The results allow to quantify the impact of climate change on long-term probabilistic structural response and highlight the need for proper modeling of the environmental parameters and climate evolution over the entire lifespan of aging structures for optimal life-cycle maintenance and management of structural and infrastructural facilities.

## 1 INTRODUCTION

Civil engineering structures are designed to guarantee the functionality and resilience of the served communities and the safety of its users over a service life spanning several decades. Multiple deterioration processes may affect the original performance of materials and structures and severely impair the level of structural safety and reliability over time (Ellingwood 2005). Design practice and management strategies of structural and infrastructural facilities should therefore account for the detrimental impact of deterioration processes, taking into account multiple factors such as durability of construction materials, quality of structural detailing, type of structural systems, and environmental exposure (Biondini & Frangopol 2016, 2019).

Daily weather fluctuations and long-term change in climate conditions can influence structural deterioration as well as the intensity and frequency of natural hazards. As highlighted by the Intergovernmental Panel on Climate Change (IPCC), climate change is unequivocal and the non-stationarity of climate and weather conditions is substantially connected to greenhouse emissions (IPCC 2021). It is hence important to investigate the impact of climate change effects on structural systems exposed to aging and deterioration, including bridges and infrastructure facilities (Nasr et al. 2020). To support the decision making process of new and existing assets, incorporating climate factors in structural performance indicators is critical to inform design, assessment, maintenance, retrofit, and repair in a life-cycle perspective (Capacci & Biondini 2019).

This paper presents a probabilistic framework for life-cycle assessment of aging structural systems exposed to environmental aggressiveness, with emphasis on Reinforced Concrete (RC) structures under corrosion processes (Biondini et al. 2004, Biondini & Vergani 2015). Climate change effects are incorporated in the proposed framework considering the long-term evolution of climate parameters affecting the deterioration rate, such as temperature and relative humidity.

The methodology is applied to a RC bridge pier subjected to chloride-induced corrosion. The results allow quantifying the detrimental impact of climate change on corrosion damage and related effects on lifetime structural safety and reliability of aging RC bridge structures.

## 2 LIFE-CYCLE STRUCTURAL ASSESSMENT IN A CHANGING CLIMATE

### 2.1 *Climate change scenarios*

The future climate is predicted through the definition of emission scenarios in relation to the evolution of population, economy, technology, energy, land use and agriculture, and described by different reference scenarios, such as the four Representative Concentration Pathways (RCPs) (Van Vuuren et al. 2011). RCP 2.6 is a mitigation scenario, RCP 4.5 and RCP 6.0 are medium stabilization scenarios and RCP 8.5 is a very high baseline emission scenario. These RCPs are representative of a larger set of scenarios reported in the literature and their denomination is associated to the value of Radiative Forcing predicted in 2100. The choice of an even number of scenarios is linked to the decision of avoiding a middle RCP, providing a manageable set of scenarios able to describe also extreme projections. RPCs have been recently integrated with the Shared Socioeconomic Pathways (SSPs) in order to characterize a range of uncertainties in the mitigation efforts required to achieve particular climate outcomes and in the adaptation efforts that could be taken to cope with the climate change effects (O'Neill et al. 2016).

Predicting spatially and temporally dependent environmental parameters considering different scenarios is a complex task. Indeed, the future climate projection depends on the selected Global Circulation Model (GCM), which is subjected to several sources of uncertainty classified in internal, model and scenario uncertainty (Madsen 2013). The inner variability of the climate systems is referred as internal uncertainty, while model uncertainty is related to the assumption and simplification considered for the definition of the GCM. Lastly, scenario uncertainty is associated with the assumptions implemented in the definition of the different climate scenarios. These uncertainties evolve in time depending on the selected environmental parameter projection and the spatial scale considered. Different analyses should be applied in order to deal with these uncertainties for the assessment of climate change impact on civil engineering structures, e.g. using different trajectories from the same GCM and considering different scenarios.

The spatial resolution of GCMs is usually ranging from tens to hundreds of kilometers while civil engineering problems often require the evaluation at the local scale. For this purpose, downscaling techniques are used in order to retrieve environmental projections for specific location. Advantages and disadvantages of the most common techniques, i.e. statistical and dynamics downscaling, are discussed in Trzaska & Schnarr (2014).

### 2.2 *Conceptual framework for life-cycle structural assessment*

The proposed methodology for life-cycle assessment of aging structures under climate change is illustrated in Figure 1. The structure of the flowchart comprises three constitutive dimensions: climate scenario modeling, mechanical damage assessment, and structural performance analysis.

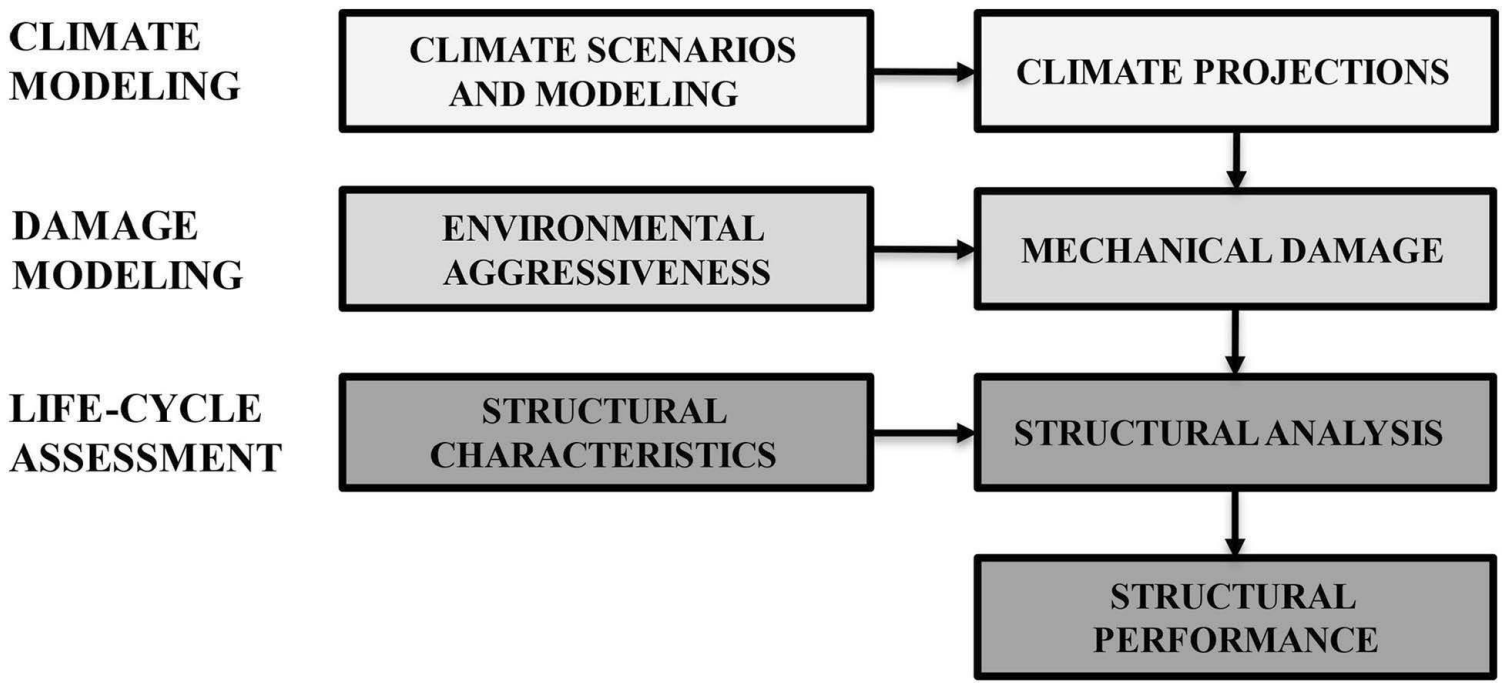

Figure 1. Methodology flowchart for performance assessment of aging structures in a changing climate.

Climate modeling provides the projections of relevant environmental parameters affecting the deterioration process based on the investigated climate scenario (e.g., RCP selection) and the input data required by the selected predictive model for climate change (e.g., site, timeframe). The projections retrieved in this phase concern the climate parameters that affect capacity deterioration and hazard rate of the investigated structural problem (e.g., temperature, relative humidity, wind speed). Then, mechanical damage is governed by environmental aggressiveness exacerbated by detrimental projections of climate parameters. Based on analytical or empirical formulations, this phase allows to describe onset and propagation of aging and deterioration processes (e.g., corrosion, freeze-thaw cycles, sulphate attacks). The characterization of mechanical damage informed by non-stationary climate conditions allows to establish the numerical models for structural analysis over time (e.g., sectional analysis, static and dynamic analysis). The problem can be formulated in probabilistic terms and integrated for practical implementation within a Monte Carlo simulation framework to account for uncertainties of basic modeling parameters encompassing all dimensions of the life-cycle structural assessment.

## 3 CORROSION MODELLING UNDER TIME-VARIANT ENVIRONMENTAL FACTORS

For RC structures exposed to chloride ingress, damage induced by the diffusive attack follows corrosion initiation and propagation and may involve corrosion of steel reinforcement and deterioration of concrete (Bertolini et al. 2013).

### 3.1 *Corrosion initiation*

The initiation time can be defined as the time necessary to achieve a critical chloride concentration at the reinforcing steel surface. Chloride ingress in concrete is generally modelled using the Fick's diffusion equation in the one-dimensional formulation:

$$D\frac{\partial^2 C}{\partial^2 x} = \frac{\partial C}{\partial t} \tag{1}$$

where $D$ is the concrete diffusivity and $C$ is the chloride concentration at the depth $x$ and time $t$. This simplified approach is considered in this paper. However, it is worth noting that for structural members with complex geometry and non-uniform chloride concentration exposure the diffusion process should be more accurately described by two- or three-dimensional patterns of concentration gradients (Biondini et al. 2004, Titi & Biondini 2016).

Over time, the aging process and the evolution of the environmental parameters, such as temperature and relative humidity, may affect the concrete diffusivity. Experimental studies show that the diffusion coefficient increases with temperature (Stewart et al. 2011). Moreover, chloride penetration occurs in the presence of water in the capillary pores. Indeed, a small relative humidity hinders chloride penetrations (Saetta et al. 1993). In order to account for non-stationary climatic conditions, the diffusion coefficient at time $t$ is evaluated as follows:

$$D(t) = D_{ref}\, f_T(t) f_{RH}(t) f_a(t) \tag{2}$$

where $D_{ref}$ is the reference diffusion coefficient for $T$=20 °C and $RH=75\%$, $f_T$ accounts for the variation of temperature, $f_{RH}$ accounts for the variation of relative humidity, and $f_a$ is an aging function.

The function $f_T$ is evaluated as follows based on the Arrhenius law (Peng & Stewart 2016):

$$f_T(t) = e^{\frac{E}{R}\left(\frac{1}{293.15} - \frac{1}{273.15 + T(t)}\right)} \tag{3}$$

where $E$ is the activation energy of the chlorides diffusion process, $R$ is the gas constant ($8.314\times10^{-3}$ kJ/mol), and $T$ is the temperature in Celsius degrees. The function $f_{RH}$ is defined based on following empirical model (Saetta et al. 1993):

$$f_{RH}(t) = \left[1 + \frac{(1 - RH(t))^4}{(1 - 0.75)^4}\right]^{-1} \tag{4}$$

where $RH$ is the concrete pore relative humidity. Finally, the function $f_a$ is evaluated as follows:

$$f_a(t) = \left(\frac{0.0767}{t}\right)^a \tag{5}$$

where $a$ is an aging coefficient.

### 3.2 *Corrosion propagation*

Corrosion initiates with the attainment of a critical concentration threshold $C_{cr}$. The propagation phase is described based on the Faraday's law in terms of corrosion penetration rate (mm/year):

$$\frac{\partial p_s}{dt} = 0.0116\ R_f\ i_{corr} \tag{6}$$

with $p_s$ = corrosion penetration depth (mm), $i_{corr}$ = corrosion electric current density (μA/cm$^2$) and $R_f$ = pitting factor, i.e. the ratio between the maximum depth measured at pit and the average penetration calculated indirectly from the weight loss of the steel bar (Stewart & Al-Harthy 2008).

Corrosion propagation may be significantly affected by the variation of environmental parameters. Indeed, a gradual reduction in the concrete resistivity along with an increase in temperature leads to increase the corrosion rate (Stewart et al. 2011). In addition, the corrosion rate is low for relative humidity values lower than 50%, it increases exponentially for relative humidity ranging from 50% to 65%, it remains nearly constant for relative humidity from 65% to 85%, and then tends to decrease (Guo et al. 2015). Moreover, the corrosion rate increases over time due to the increase of chloride concentration (Liu & Weyers 1998) and decreases due to the formation of corrosion products (Vu & Stewart 2000). Based on these results, the influence of environmental parameters on the time-variant corrosion penetration rate is investigated in this paper based on the following relationship (Casti 2021, Nava 2022):

$$i_{corr}(t) = i_{corr,ref}\varphi_{RP}(t)\varphi_T(t)\varphi_{RH}(t)\varphi_C(t) \tag{7}$$

where $i_{corr,ref}$ is the reference corrosion rate for $T$=20 °C and $RH$=75%, and the time-variant functions $\varphi_{RP}(t)$, $\varphi_T(t)$, $\varphi_{RH}(t)$, and $\varphi_C(t)$ account for the effects associated with formation of rust products and variation of temperature, relative humidity, and chloride concentration, respectively. The function $\varphi_{RP}$ is described as follows (Vu & Stewart 2000):

$$\varphi_{RP}(t) = 0.85t_p^{-0.29} \tag{8}$$

where $t_p$ is the corrosion initiation time. The function $\varphi_T$ is based on the following stepwise linear relationship (Stewart et al. 2011):

$$\varphi_T(t) = [1 + K_T(T(t) - 20)] \tag{9}$$

where $K_T$=0.025 or $K_T$=0.073 for temperature $T \leq 20°C$ and $T > 20°C$, respectively. Finally, the functions $\varphi_{RH}$ and $\varphi_C$ are based on the following formulations (Guo et al. 2015):

$$\varphi_{RH}(t) = e^{-6000(RH(t)-0.75)^6} \tag{10}$$

$$\varphi_C(t) = \frac{C(t) + C_{cr}}{2\ C_{cr}} \tag{11}$$

Sensitivity analyses on the proposed corrosion model are reported in Nava (2022).

### 3.3 *Mechanical damage modeling*

Corrosion damage is modeled considering the reduction in the mass of steel reinforcement, concrete strength, and steel ductility as proposed in Biondini & Vergani (2015). The reduction of cross-sectional area $A_s$ of steel bars is the most relevant damage induced by corrosion:

$$A_s(t) = [1 - \ \delta_s(t)]A_{s0} \tag{12}$$

where $A_{s0}$ is the area of the undamaged bar and $\delta_s$ is a damage function of the corrosion penetration depth related to the type of corrosion, i.e. uniform or localized (pitting) corrosion. Delamination and spalling are related to the propagation of longitudinal cracks along the concrete cover due to the formation of oxidation products on the steel bars. The effects of such degradation process can be modelled by reducing the effective concrete compression strength $f_c$ in terms of the damage indices evolution of the steel longitudinal reinforcement. Furthermore, reduction of steel ductility is experimentally reported to be related to percentage of steel mass loss induced by corrosion to brittle failure. This effect can be modelled reducing the ultimate strain capacity $\varepsilon_{su}$ of the steel bars.

## 4 APPLICATION

### 4.1 *RC bridge pier*

The proposed methodology is applied to life-cycle reliability assessment of the RC two-column bridge pier shown in Figure 2. Each column has hollow-core rectangular cross-section with overall dimensions 3.8 m × 3.4 m (Figure 2a). The steel reinforcement varies over the length of the pier with a concrete cover of 75 mm. The bending capacity of the RC cross-sections A-A′ (Figure 2b) and B-B′ (Figure 2c), with overall reinforcement $A_s$=1491 cm$^2$ and $A'_s$=745 cm$^2$ for cross-section A-A′ and $A_s$=718 cm$^2$ and $A'_s$=583 cm$^2$ for cross-section B-B′, is investigated. An axial force $N$=67.09 MN associated with dead loads is considered. The cross-sections are modelled based on a fiber approach in OpenSees (Mazzoni et al. 2006). The constitutive model for concrete (Concrete04) accounts for confinement (Figure 2d). The stress-strain law for steel bars is elastic-plastic with strain hardening (Steel01). The bridge pier is assumed to be exposed to chloride attack over internal and external surfaces with uniform concentration $C_s$.

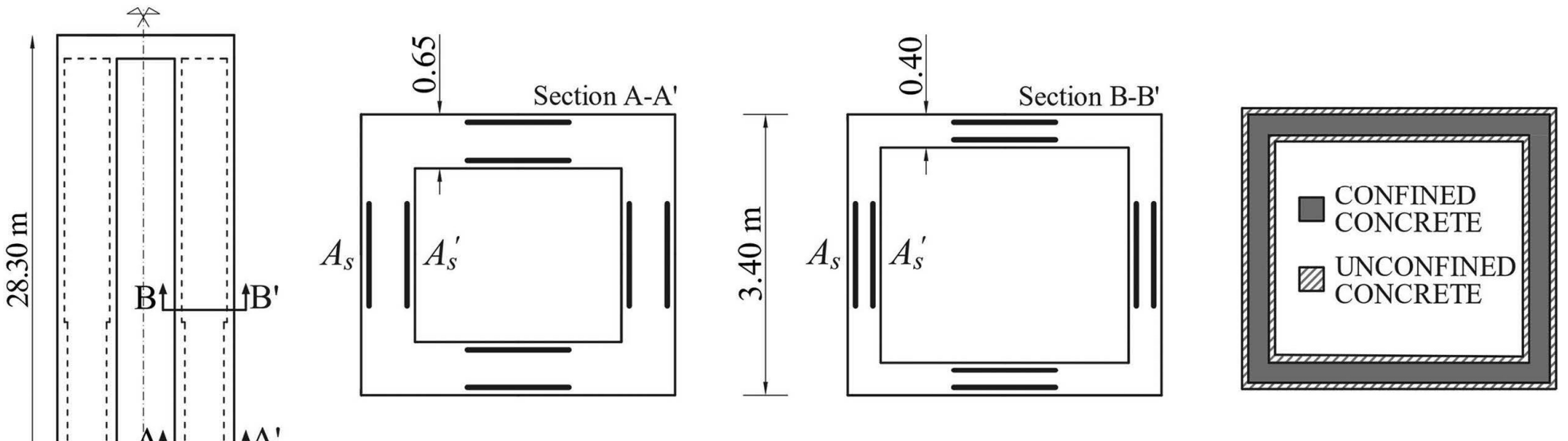

Figure 2. (a) RC bridge pier; (b) cross-section A-A′; (c) cross-section B-B′; (d) concrete confinement.

The analysis timeframe ranges from the year 2000 to the year 2100. The three climate scenarios shown in Figure 3 are considered: (1) very high baseline emission scenario (RCP 8.5); (2) mitigation scenario (RCP 2.6); and (3) No Climate Change (NCC). Temperature projections in Figure 3a are based on the fourth version of the Community Climate System Model (Gent et al. 2011). Projections are downscaled with respect to the station located in Viale Marche, Milan, Italy (Casti 2021) and the analyses are performed considering annual time scale for scenario NCC and monthly time scale for scenarios RCP 2.6 and RCP 8.5, by taking into account also cyclic temperature variations. Linear projections of relative humidity as shown in Figure 3b are considered based on Bastidas-Arteaga et al. (2013).

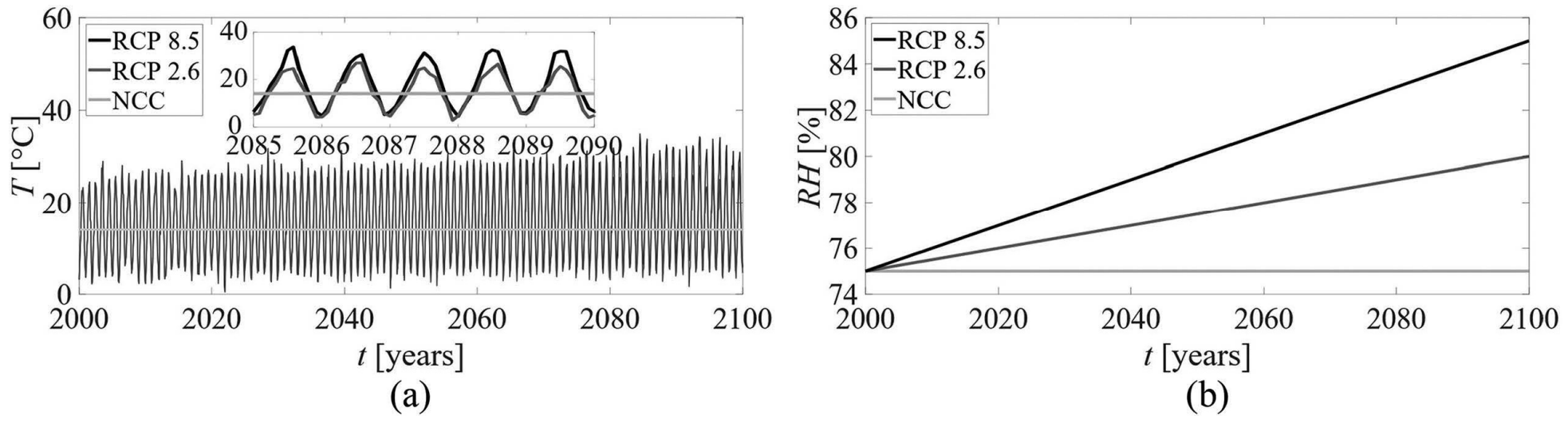

Figure 3. Temperature and relative humidity projections under different climate scenarios.

The probabilistic analysis accounts for uncertainties in mechanical and environmental damage stressors, including surface chloride concentration $C_s$, critical chloride concentration $C_{cr}$, reference diffusivity coefficient $D_{ref}$, aging coefficient $a$, reference current density $i_{corr,ref}$, pitting factor $R_f$, concrete strength $f_c$, and steel strength $f_y$. Probability distribution types, probabilistic descriptors and bounds of the random variables are listed in Table 1.

Table 1. Random variables, distribution types, and descriptors. Adapted from multiple sources (Nava 2022).

| Random Variable | Distribution Type | Mean $\mu$ | C.o.V | Bounds |
|---|---|---|---|---|
| Surface chloride concentration $C_s$ [%wt./c] | Truncated Normal | 3.0 | 0.30 | $C_s \geq 0$ |
| Critical chloride concentrations $C_{cr}$ [%wt./c] | Beta | 0.60 | $0.15/\mu$ | $0.2 \leq C_{cr} \leq 2$ |
| Reference diffusivity $D_{ref}$ [$10^{-12}$ m$^2$/s] | Truncated Normal | 15.8 | 0.20 | $D_c \geq 0$ |
| Aging coefficient $a$ [-] | Beta | 0.30 | $0.12/\mu$ | $0 \leq a \leq 1$ |
| Reference current density $i_{corr,ref}$ [μA/cm$^2$] | Lognormal | 2.586 | 0.67 | $i_{corr} \geq 0$ |
| Pitting factor $R_f$ [-] | Gumbel | 7.1 | $0.17/\mu$ | $R_f \geq 0$ |
| Unconfined concrete strength $f_c$ [MPa] | Lognormal | 40.02 | $5/\mu$ | $f_c \geq 0$ |
| Steel yielding strength $f_y$ [MPa] | Lognormal | 487.58 | $30/\mu$ | $f_y \geq 0$ |

### 4.2 *Nominal structural analysis*

Deterministic structural analyses are carried out for the two investigated cross-sections by assuming the mean values of the random variables as nominal values.

The bending moment $M$ vs curvature $\chi$ diagrams in the vertical plane normal to the bridge axis (strong axis of the cross-sections) are shown in Figure 4a at the years 2000 (light gray line) and 2100 considering scenarios NCC (dark gray line) and RCP 8.5 (black line). Limit states are also reported over the capacity curves for concrete first cracking (star marker), concrete cover spalling (circle marker), and ultimate curvature (square marker). At the time 2100, cover spalling is not reported since it occurs earlier for corrosion products expansion. The results highlight that the effect of climate change significantly reduces the lifetime structural capacity of the bridge pier.

Figure 4b and Figure 4c show the axial force $N$ vs bending moment $M$ capacity curves associated with cracking and peak bending capacity limit states, respectively for cross-section A-A′ (Figure 4b) and cross-section B-B′ (Figure 4c). These results show that the detrimental impact of climate change on the cracking limit state is exacerbated by the axial force level, while the effects on the peak bending capacity are almost independent on the axial force. Additional results related to multiple limit states of the cross-sections are available in Nava (2022).

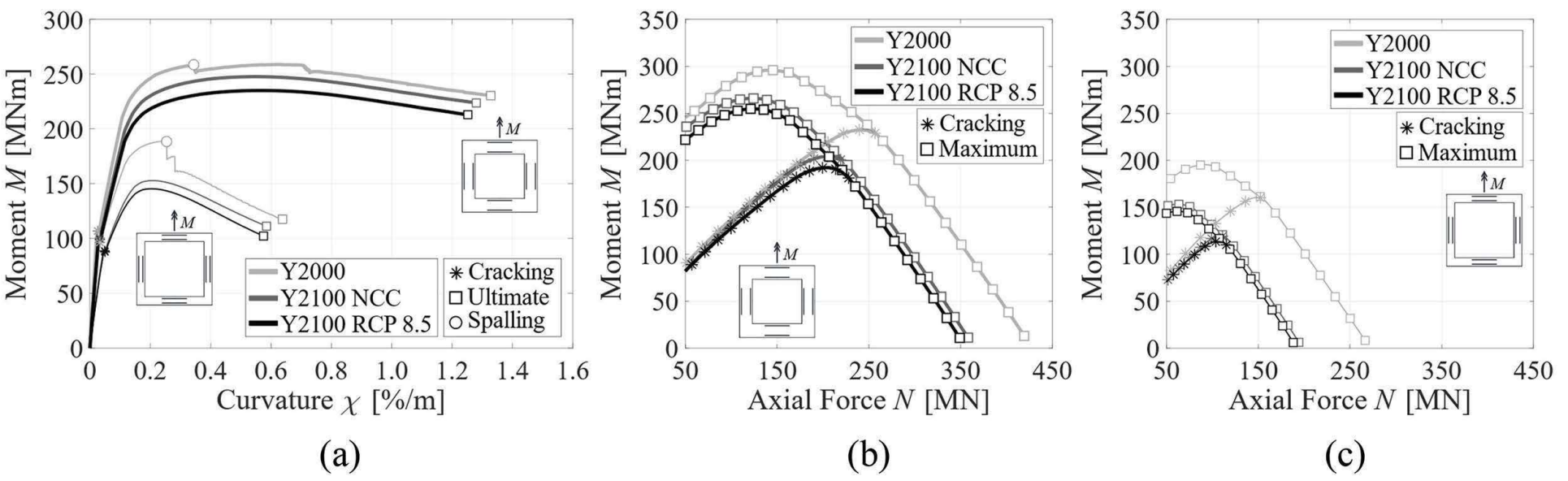

Figure 4. (a) M-χ diagrams; *M-N* capacity curves of (b) cross-section A-A′ and (c) cross-section B-B′.

## 4.3 *Life-cycle structural reliability analysis*

The results of a Monte Carlo simulation from a sample of $10^5$ realizations are shown in Figure 5 in terms of bending moment $M$ vs curvature $\chi$ diagrams of cross-section A-A′. Figure 5a compares median, 16% and 84% fractiles at years 2000 and 2100 under no climate change (NCC) and RCP 8.5. The results for the entire samples are also shown by strong transparency in Figure 5b (Year 2000) and Figure 5c (Year 2100 under RCP 8.5) to better highlight relevant statistical features of the $M$-$\chi$ diagrams, such as centrality, variability, and skewness.

Life-cycle structural reliability is assessed with a 10-year time step considering as limit state the attainment of a target peak bending strength $M_t$=166 MN. The sample size is adaptively selected to mitigate the computational effort and guarantee a sufficient level of estimate accuracy. The results in terms of time-variant reliability index $\beta(t)$ are reported in Figure 6a. The detrimental impact of climate change can be appreciated at the year 2100 by a larger reliability index decay, approximately of 20% for RCP 2.6 and 30% for RCP 8.5, with respect to NCC scenario. These results confirm the relevant effects of climate change on life-cycle structural reliability.

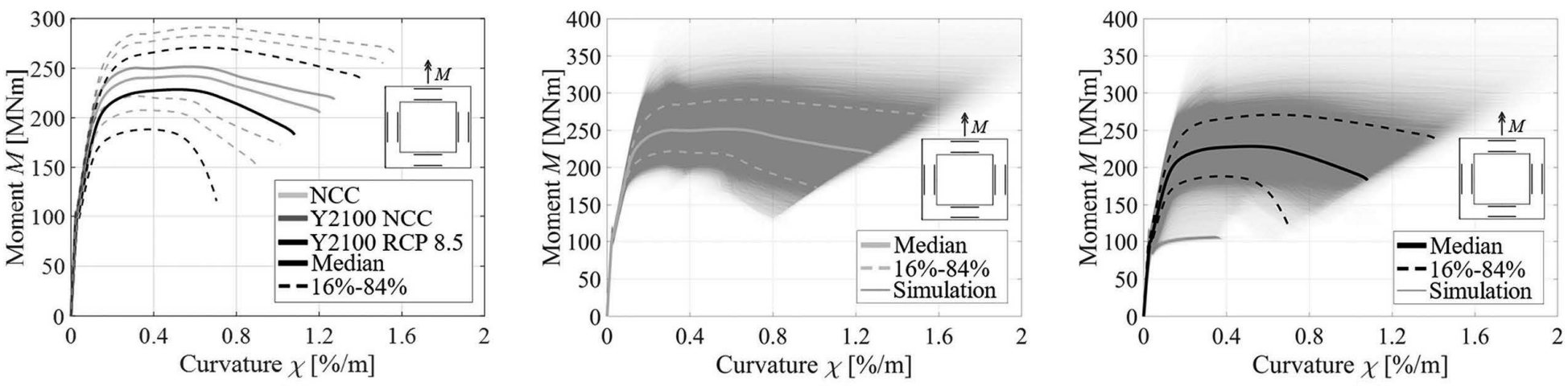

Figure 5. $M$-$\chi$ diagrams of cross-section A-A′: (a) Median, 16% and 84% fractiles at years 2000 and 2100 under no climate change (NCC) and RCP 8.5; Results for the entire samples of $10^5$ Monte Carlo realizations at years (b) 2000 and (c) 2100 under RCP 8.5.

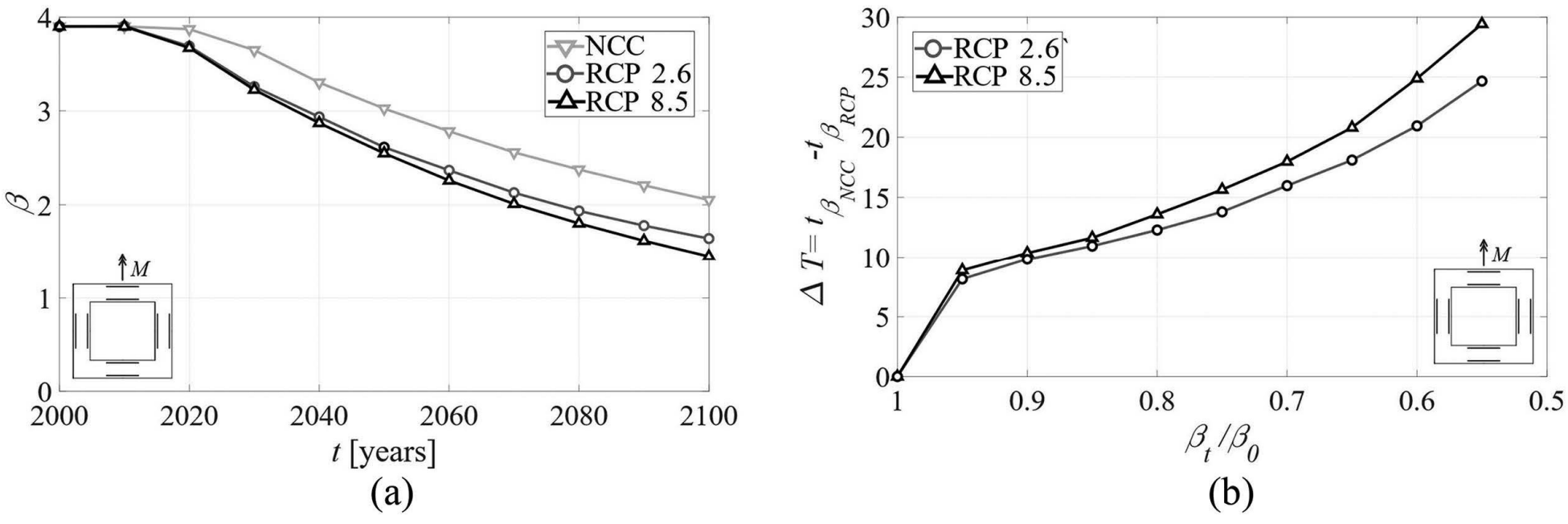

Figure 6. Life-cycle reliability analysis of cross-section A-A′ under no climate change (NCC) and two climate change scenarios (RCP 2.6 and RCP 8.5): (a) Time-variant reliability index $\beta(t)$; (b) Time lag $\Delta T$ between NCC and RCP scenarios vs target reliability index $\beta_t$.

An effective quantitative measure of the impact of a changing climate can also be achieved by computing the time lag associated with the attainment of a target reliability index $\beta_t$ between two scenarios with and without climate change effects. Figure 6b shows the time lag $\Delta T$ vs the target reliability index $\beta_t$ normalized with respect to the initial reliability index $\beta_0$ (i.e., $\beta_t/\beta_0$). As an example, for $\beta_t$=0.55$\beta_0$ the time lag is approximately 30 years for the RCP 8.5 scenario and 25 years for the RCP 2.6 scenario. It is worth noting that the relevant time lag associated to the mitigation scenario RCP 2.6 is mainly due to the effects of the cyclic variation of temperature. Indeed, neglecting it, as highlighted in additional analyses reported in Nava (2022), the detrimental effects of climate change are reduced.

## 5 CONCLUSIONS

The effects of non-stationary climate conditions on life-cycle performance and reliability of RC structures under corrosion has been investigated. The deterioration model incorporates future projections of climate parameters leading to increase damage caused by corrosion. The results obtained for a RC bridge pier cross-section allowed to quantify the impact of climate change on long-term structural reliability and highlighted the importance of modeling degradation taking environmental exposure and climate evolution into account. Neglecting seasonal fluctuations and climate trends may lead to an underestimation of the reliability index over time and, in turn, an overestimation of times for inspections, maintenance, retrofit, and repair.

This paper focused on the impact of climate change on the life-cycle reliability of individual concrete structures under corrosion. Life-cycle maintenance and management of resilient infrastructure systems should also be investigated accounting for climate change effects on the interaction of an increasingly aggressive environment with multiple hazards and considering the harmful impact of loss of functionality in critical system components. Further research is hence necessary to incorporate the proposed framework into risk-based procedures for optimal mitigation and adaptation strategies with a systemic perspective.